\documentclass[oneside,onecolumn,superscriptaddress,showpacs]{revtex4}
\usepackage{epsf,amssymb}
\usepackage{graphicx}
\begin{document}
\bibliographystyle{plain}
\nocite{*}
\title{Investigation of the fundamental constants stability based on the
reactor Oklo burn-up analysis}
\author{M.S. Onegin}
\email{onegin@thd.pnpi.spb.ru} \affiliation{Petersburg Nuclear Physics Institute, Gatchina, Leningrad district, Russia, 188 300}
\date{December 29, 2014}
\begin{abstract}
New severe constraints on the variation of the fine structure constant have been obtained from reactor Oklo analysis in our previous work. We investigate here how these constraints confine the
parameter of BSBM model of varying $\alpha$.  Integrating the coupled system of equations from the Big Bang up to the present time and taking into account the Oklo limits we have obtained the
following margin on the combination of the parameters of BSBM model:
$$ |\zeta_m (\frac{l}{l_{pl}})^2|<6\cdot 10^{-7}, $$
where $l_{pl}=(\frac{G\hbar}{c^3})^{\frac{1}{2}} \approx 1.6 \cdot 10^{-33}$ cm is a Plank length and $l$ is the characteristic length of the BSBM model. The natural value of the parameter
$\zeta_m$ - the fraction of electromagnetic energy in matter - is about $10^{-4}$. As a result it is followed from our analysis that the characteristic length $l$ of BSBM theory should be
considerably smaller than the Plank length to fulfill the Oklo constraints on $\alpha$ variation.
\end{abstract}



\maketitle

\section{Introduction}

The confirmation of the temporal variation of the fundamental constants would be the first indication of  the universe expansion influence on the micro physics~\cite{Review}.

Shlyakhter was the first who showed that the variation of the fundamental constants could lead to measurable consequences on the Sm isotops concentrations in the ancient reactor
waste~\cite{Shlyakhter}. Later Damur and Dyson~\cite{Damour:1996zw} for Zones 2 and 5 and also Fujii~\cite{Fujii:1998kn} for Zone 10 of reactor Oklo made more realistic analysis of the possible
shift of fundamental constants during the last $~2\cdot 10^9$ years based on the isotope concentrations in the rock samples of Oklo core. In this investigation the idealized Maxwell spectrum of
neutrons in the core was used. The efforts to take into account more realistic spectrum of neutrons in the core were  made in works~\cite{Petrov:2005pu,Lamur06}. New severe constraints on the
variation of the fine structure constant have been obtained from reactor Oklo analysis in work \cite{oneg1}:
$$-0.7\cdot 10^{-8} < \delta \alpha/\alpha < 1.0\cdot 10^{-8} $$

We investigate here how these constraints confine the parameter of BSBM model \cite{SBM} of varying $\alpha$. This theory combines Bekenstein extension of electrodynamics~\cite{Bekenstein} with
varying alpha to include gravitational effects of new scalar field $\psi$. It respects covariance, gauge invariance, causality and has only two free parameters: the fraction of electromagnetic
energy $\zeta_m$ in the total energy of matter including dark matter as well as the dimensional parameter $l$ which is having sense of characteristic length. As a result of our analysis we get
the constraints on the combination of the parameters of BSBM model.

\section{BSBM theory}

BSBM theory~\cite{SBM} is the extension of the Bekenstein~\cite{Bekenstein} theory to include dynamics of the gravitational field. Total action of this theory has a form:
\begin{equation}
S=\int d^4x \sqrt{-g}(L_g+L_{mat}+L_{\psi}+e^{-2 \psi} L_{em})
\end{equation}
where $L_{\psi}=-\frac{\omega}{2} \partial_{\mu}\psi \partial^{\mu}\psi$ and $L_{em}=-\frac{1}{4}f_{\mu \nu}f^{\mu \nu}$. A parameter $\omega$ here is definite as $\omega = \frac{\hbar c}{l^2}$
where dimensional parameter $l$ is having sense of characteristic length. Fine structure constant expressed via $\psi$ with the equation: $\alpha = \frac{e_0^2}{\hbar c} e^{2 \psi}$. Varying
$\psi$ we get the following equation:
\begin{equation}
\Box \psi = \frac{2}{\omega} e^{-2 \psi} L_{em}.
\end{equation}
For pure radiation $L_{em}=(E^2-B^2)/2=0$, so $\psi$ remains constant during radiation domination epoch. Only in matter domination epoch changes in $\alpha$ take place. The only contribution to
variation of $\psi$ come mainly from pure electrostatic or magnetostatic energy. It is convenient to work in the following parameter: $\zeta _N = m_N^{-1} \langle N \vert L_{em} \vert N \rangle$
and according to~\cite{Gasser} $\zeta_p=-0.0007$ and $\zeta_n=0.00015$. Varying the metric tensor and using Friedmann metric we get the following Friedmann equation:
\begin{equation}
\left( \frac{\dot{a}}{a} \right)^2=\frac{8\pi G}{3}\left[ \varrho_m (1+\zeta_m e^{-2\psi}) + \varrho_r e^{-2\psi} + \frac{\omega}{2} \dot{\psi}^2 + \Lambda/8\pi \right],
\end{equation}
and the equation for $\psi$ takes form:
\begin{equation}
\ddot{\psi} + 3 H \dot{\psi} = - \frac{2}{\omega} e^{-2\psi} \zeta_m \varrho_m,
\end{equation}
where $H=\frac{\dot{a}}{a}$.

We have also energy conservation equations:
\begin{eqnarray} & \dot{\varrho}_m + 3H \varrho_m = 0, & \nonumber \\
 & \quad \dot{\varrho}_r + 4H \varrho_r = 2 \dot{\psi} \varrho_r, & \nonumber \end{eqnarray} which have solutions: $\varrho_m \sim a^{-3} $, and $e^{-2\psi}\varrho_r \sim a^{-4}.$

Let use critical density: $$\varrho_c(t)=\frac{3H^2(t)}{8\pi G},$$ and use also the fractions of all densities relative to critical: $\Omega_m, \Omega_r, \Omega_{\Lambda}$. Index $(0)$ will
denote the present values of these fractions. We use the ordinary values for these fractions at present: $\Omega_m^{(0)}=0.3$, $\Omega_r^{(0)}=2.0\cdot 10^{-5}$, and $\Omega_{\Lambda}$ is
determined from the condition that the Universe is flat.

Then the Friedmann equation takes form:
\begin{equation}
\left( \frac{\dot{a}}{a} \right)^2=H_{(0)}^2 \left[ \Omega_m^{(0)} \left( \frac{a_0}{a} \right)^3 (1+\zeta_m e^{-2\psi}) + \Omega_r^{(0)} \left( \frac{a_0}{a} \right)^4 e^{-2\psi_0} +
\Omega_{\Lambda}^{(0)}+ \frac{\omega}{2} \frac{ \dot{\psi}^2}{\varrho_{c0}} \right],
\end{equation}
and equation for $\psi$:
\begin{equation}
\frac{d}{dt}(a^3\dot{\psi})=N e^{-2\psi}.
\end{equation}
Here constant $N$ is equal to $N=-2\frac{\zeta_m}{\omega}\Omega_m^{(0)}\varrho_{c0}$. For negative $\zeta_m$ this constant is positive and has the following dependence on the ratio of
characteristic and Plank lengthes:
\begin{equation}
N \sim \zeta_m \left( \frac{l}{l_{pl}} \right)^2.
\end{equation}

The result of the numerical integration of these equations is presented of Fig.1 for the variation of different components of energy density with red shift $z$, and on Fig.2 for the variation of
fine structure constant $\alpha$. Here we use the notation: $\Omega_{\psi} = \frac{\omega}{2} \frac{ \dot{\psi}^2}{\varrho_{c0}}$. We took the value of the characteristic length $l$ equal to
$l_{pl}$ during this analysis, and assigned the following value for the the parameter $\zeta_m$:
$$ \zeta_m = \frac{\Omega_b^{(0)}}{\Omega_m^{(0)}}\zeta_p \approx - 10^{-4}, $$
where $\Omega_b^{(0)}=0.03$ - is the fraction of energy density in the Universe due to ordinary baryonic matter. The initial values of the scalar field $\psi$ for the second order differential
equation (6):  the value of the scalar field $\psi$ and its derivative during the radiation epoch was taken  in such a manner that the present value of the fine structure constant coincide with
experiment, and it appeared that the initial value of the $\dot{\psi}$ during the radiation domination epoch could be assigned a rather arbitrary value because the result of integration
influenced rather weakly by this choice.

\begin{figure}
\centerline{ \includegraphics[width=12cm]{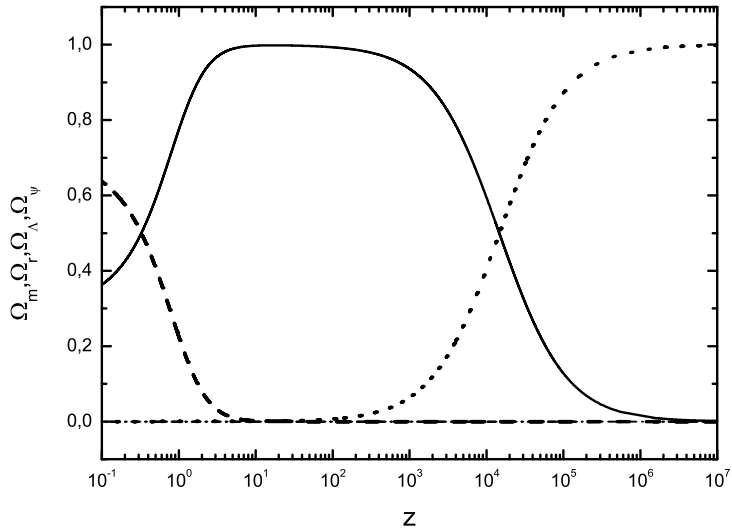}}\caption{Red shift dependence of the different components of the energy density. Solid line - matter component, dot line - radiation
component, dash line - cosmological constant and $\Omega_{\psi}$ - dash-dot line.} \label{dens}
\end{figure}

\begin{figure}
\centerline{ \includegraphics[width=12cm]{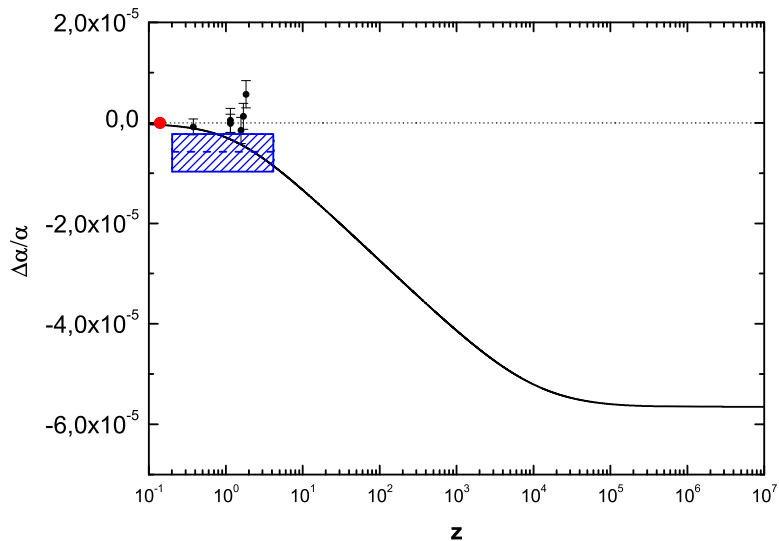}}\caption{Red shift dependence of the fine structure constant. Dashed box - 3 $\sigma$ experimental results for Keck telescope~\cite{Webb},
closed circles - experimental results from VLT telescope (data were taken from work~\cite{VLT}), red circle at $z=0.14$ - Oklo result.} \label{alpha}
\end{figure}

As it is followed from Figure~\ref{dens}, the scalar field $\psi$ influence rather weakly on the variation of the different components of the energy density with red shift. The total variation
of alpha during the whole history of the Universe is about $6\cdot 10^{-5}$ (as is followed from Figure~\ref{alpha}) which is not contradict Big Bang and radiation recombination
constraints~\cite{Review}. On the other side the Oklo analysis predict about zero result for $\Delta \alpha / \alpha$ with the experimental  error which could be seen in Figure~\ref{alpha}) if
we increase the scale of Figure~\ref{alpha} one hundred times. We investigate the constraints  on the parameters of BSBM model followed from Oklo analysis in the next section.

\section{Constraints based on Oklo analysis on parameters of BSBM model }

 In analysis of Oklo data~\cite{oneg1} we obtained the following constraints on the variation of the fine structure constant
\begin{equation}
\label{varalpha} -0.7\cdot 10^{-8} \le \frac{\Delta \alpha}{\alpha} \le 1.0\cdot 10^{-8}
\end{equation}
during the past $2\cdot 10^9$ years. The age of the reactor $1.93\cdot 10^9$ years corresponds to red shift parameter $z=0.14$. We use here also previous constraints obtained
in~\cite{Petrov:2005pu}:
\begin{equation}
\label{varalpha} -5.8\cdot 10^{-8} \le \frac{\Delta \alpha}{\alpha} \le 6.6\cdot 10^{-8}, \nonumber
\end{equation}
and in~\cite{Lamur06}:
\begin{equation}
\label{varalpha} -1.1\cdot 10^{-8} \le \frac{\Delta \alpha}{\alpha} \le 2.4\cdot 10^{-8}.
\end{equation}

\begin{figure}
\centerline{ \includegraphics[width=12cm]{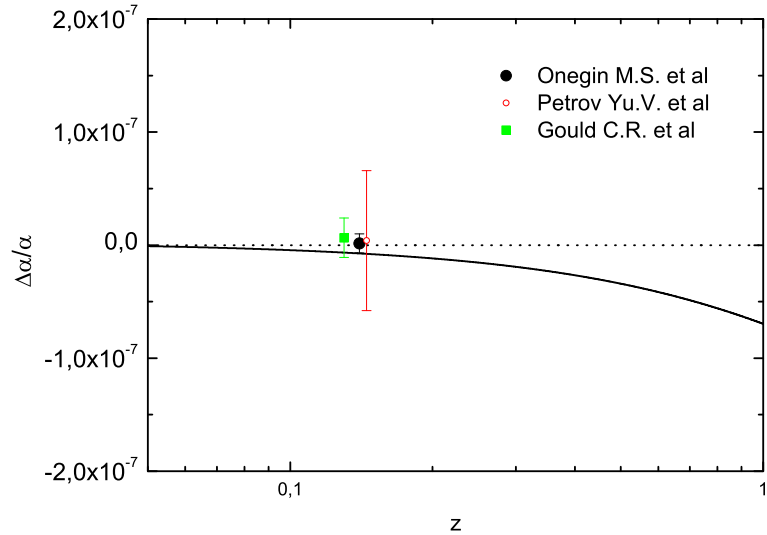}}\caption{Red shift dependence of the fine structure constant. Closed circle - result from work~\cite{oneg1}, open circle -
result~\cite{Petrov:2005pu}, box - result~\cite{Lamur06}.} \label{Oklo}
\end{figure}

All these constraints are shown on Figure~\ref{Oklo}. To provide the solution of the equations (5) and (6) which doesn't contradict the result of work~\cite{oneg1}  (see Figure~\ref{Oklo}), we
have to set rather severe constraints on the combinations of the parameters of BSBM model. They have to satisfy the following inequality: $$ \left|\zeta_m \left( \frac{l}{l_{pl}} \right) ^2
\right| < 6\cdot 10^{-7}. $$ For realistic value $\zeta_m=-10^{-4}$ to fulfill this inequality we have to demand that:
\begin{equation}
l < 0.1 \, l_{pl}.
\end{equation}

\section{Discussion of the results}

A theoretical framework under very general assumptions was worked out by Bekenstein to admit the variation of the fine structure constant. A characteristic length  $l$    enters into it. An
experimental constraint rules out $\alpha$ variability of any kind if it is in clear conflict with predictions of the framework for   $l$    no shorter than the fundamental length $l_{pl}$
(\cite{Bekenstein}). As a result of Oklo analysis we get $l < 0.1 \, l_{pl}$ the Oklo geophysical constraints strongly rule out all $\alpha$ variability.

In this analysis we have considered only the variation of electromagnetic fine structure constant $\alpha$. If other fundamental constants also varies the picture would be more complicated as
well as the analysis of the Oklo phenomenon and the analysis of the cosmological variation of $\alpha$. To do such analysis in our opinion would be too early because till now we haven't had any
convincing manifestations of the cosmological variations of the other fundamental constants~\cite{Review}.

\section{Acknowledgments}
The author would like to express his gratitude to S. Karshenboim and M.S. Yudkevich for useful discussions and critical remarks. This work was partly supported by the RSCF grant (project
14-22-00281).


\end{document}